\documentclass[twocolumn]{webofc}
\usepackage[varg]{txfonts}   % Web of Conferences font
\usepackage{graphicx,amsmath,amssymb,bm,xcolor,tabularx}
\usepackage{hyperref}
\begin{document}
\begin{sloppypar}

\title{Comparison of Practical Expressions for E1 Photon Strength Functions }

\author{\firstname{Oleksandr} \lastname{Gorbachenko}\inst{1}%\fnsep\thanks{\email{gorbachenko@univ.kiev.ua}}
        \and
        \firstname{Igor} \lastname{Kadenko}\inst{1}%\fnsep\thanks{\email{}}
        \and
        \firstname{Vladimir} \lastname{Plujko}\inst{1,2}\fnsep\thanks{\email{plujko@gmail.com}}
        \and
        \firstname{Kateryna} \lastname{Solodovnyk}\inst{1}%\fnsep\thanks{\email{}}
        % etc.
}

\institute{Taras Shevchenko National University of Kyiv, Volodymyrska str.60, Kyiv 01601, Ukraine \and
Institute for Nuclear Research, Prosp. Nauki 47, Kyiv 03028, Ukraine}

\abstract{

The closed-form expressions for the photon strength functions (PSF) are tested using the gamma-decay data of OSLO group. The theoretical calculations are performed for the Lorentzian models of PSF for electric and magnetic dipole gamma-rays. The criteria of minimum of least-square value as well as the root-mean-square deviation factor are used. It is shown that a rather good agreement is obtained within the Simple Modified Lorentzian model for E1 PSF modelling.
} %end of abstract
%
% 21.60.-n Nuclear structure models and methods
% 24.30.Cz Giant resonances
% 25.20.-x Photonuclear reactions

\maketitle
\section{Introduction}
\label{intro}

Many nuclear physics applications, such as radiation shielding and radiation transport analyses, fission and fusion reactor technologies, nuclei medicine and radiotherapy etc., require a reliable data on photoabsorption and photodecay cross section data. The PSFs play an important role in the calculations of such nuclei data and they are an important constituent of the calculations of various nuclei properties \cite{RIPL,RIPL2,RIPL3}.

The reliability of the PSF predictions can be greatly improved by the use of microscopic or semimicroscopic approaches (\cite{Gor2019}). When compared with experimental data and considered for practical applications, all such calculations need however some phenomenological corrections. Moreover, most of the microscopic and semimicroscopic approaches assume spherical symmetries so that phenomenological corrections need to be included in order to properly describe the gamma-strength function in deformed nuclei. In additional, the results of the microscopic calculations for the whole nuclear chart can not be provided without super computers. Alternatively, the phenomenological models can be used in the deformed nuclei without time consuming calculations.

Among the phenomenological models, a Lorentz shape is preferable for approximation of the E1 PSF (see \cite{Plu2011,Plu2018} and refs. therein). The analytical models considered here are the Standard Lorentzian (SLO), the Generalized Lorentzian (GLO), the Simple Modified Lorentzian (SMLO) \cite{RIPL,RIPL2,RIPL3,Plu2011,Plu2018,Gor2019}, the Triple Lorentzian (TLO) \cite{Jun2008,Gro2017}. Note that, the calculations of E1 PSF within microscopic both quasiparticle random-phase approximation (QRPA) and shell-model (SM) were compared with the SMLO strengths in ref.\cite{Gor2019} and had been shown that a rather good agreement was obtained.

The experimental data  on $\gamma$-transitions near the neutron separation energy include large contribution of the M1 transitions \cite{Gor2018,Gor2019,Krt2004,Mum2017}. In this contribution the sum of M1 and E1 PSF is tested for description of the photodecay data by OSLO group \cite{Oslo} for even-even atomic nuclei in the range of the $\gamma$-rays till $\sim$ 10 MeV. The  data are compared with theoretical calculations using the merit functions of the least-square values and the root-mean-square (rms) deviation factor.

\section{Analytical expressions of E1 PSF for photodecay}
\label{sec:1}

For the heated nuclei, the general analytical expression for E1 PSF of photodecay $\overleftarrow{f}^{\alpha }$ can be presented in the following form \cite{RIPL,RIPL2,RIPL3}:

\begin{equation}
\label{EQ_1}
\begin{array}{c}
{\displaystyle
\overleftarrow{f}^{\alpha } (\varepsilon _{\gamma })= \Phi (\varepsilon _{\gamma }, T_{f}),} \\
{\displaystyle
\Phi (\varepsilon _{\gamma },T) = \frac{1}{3\cdot (\pi \hbar c)^{2} } \sum _{j=1}^{j_{m} }\sigma_{\rm{TRK}} s_{j}^{\alpha } \frac{{F}_{j}^{\alpha } (\varepsilon _{\gamma },T )}{\varepsilon _{\gamma } }.  }
\end{array}
\end{equation}

Here, {$\alpha $=} SLO, SMLO, GLO, TLO models, $\varepsilon _{\gamma } $ - the gamma-ray energy, $T_f$ - temperature of the final states of the nuclei during the gamma decay, index $j$ numbers the normal modes of giant vibrations of Isovector Giant Dipole Resonance (GDR): $j_{m} =1$ for spherical nuclei, $j_{m} =2$ for  axially symmetric ones, and $j_{m} =3$ for nuclei with triaxial shape (TLO model); factor $s_{j}^{\alpha } $ is a weight of the \textit{j}-mode;
$\sigma_{\rm{TRK}}$ is the Thomas-Reiche-Kuhn (TRK) sum rule $\sigma _{\rm{TRK}} =15A(1-I^{2} ) \ \ {\rm mb}\cdot {\rm MeV}$ with $I=(N-Z)/A$ for the neutron-proton asymmetry factor. A weight of the \textit{j}-mode determines cross section $\sigma _{r,j}^{\alpha } =(2/\pi )\sigma _{\rm{TRK}} \cdot s_{j}^{\alpha } /\Gamma _{r,j}^{\alpha } $ of $j$ -mode at GDR resonance energy $E_{r,j}^{\alpha } $, where $\Gamma _{r,j}^{\alpha }$ width of $j$-resonance mode.

The line-shape funciton for gamma-decay for all models (except GLO) can be given as
\begin{equation}
\label{EQ_2}
\begin{array}{l}
{\displaystyle {{F}_{j}^{\alpha } (\varepsilon _{\gamma }, T)=} } \\
{\displaystyle = {L}^{\alpha }(\varepsilon _{\gamma } ,T)\cdot \frac{2}{\pi } \frac{\varepsilon _{\gamma }^{2} \, \Gamma _{j}^{\alpha }(\varepsilon _{\gamma }, T) }{\left({\varepsilon _{\gamma }^{2} -(E_{r,j}^{\alpha } )^{2}} \right)^{2} +\left({\Gamma _{j} ^{\alpha }(\varepsilon _{\gamma }, T) \cdot \varepsilon _{\gamma } }\right)^{2}},}
\end{array}
\end{equation}

\noindent where $ L^{\alpha }(\varepsilon _{\gamma},T) = 1 $ for the SLO and TLO models and

\begin{equation}
\label{EQ_3}
{\displaystyle  L^{SMLO }(\varepsilon _{\gamma},T)= {1}/ \left({1-\exp (-\varepsilon _{\gamma } /T)} \right)}
\end{equation}

\noindent for the SMLO approach. It determines the enhancement of the radiative strength function in a heated nucleus as compared to a cold nucleus {\cite{Plu1990,RIPL2,RIPL3,Gor2019}}, and can be interpreted as the average number of 1p--1h states excited by an electromagnetic field.

The function of GLO model consists of two components: a Lorentzian ${F}_{j}^{{\rm GLO}}$  defined by eq.({\ref{EQ_2}}) with $ L^{\alpha }(\varepsilon _{\gamma},T) = 1 $ and additional term within Fermi-liquid theory \cite{Kad1983} i.e. ${F}_{j}^{{\rm GLO}}$ in eq.({\ref{EQ_1}}) can be replaced  on
\begin{equation}
\label{EQ_4}
\begin{array}{l}
{\displaystyle F_{j}^{{\rm GLO}} (\varepsilon _{\gamma }, T)} %\\
{\displaystyle + \frac{\pi }{2} \cdot \varepsilon _{\gamma } \cdot \frac{0.7\, \Gamma _{j}^{{\rm GLO}} (\varepsilon _{\gamma } =0,T)}{(E_{r,j}^{{\rm GLO}} )^{3} }.}
\end{array}
\end{equation}

The shape width in heated nuclei is the temperature-dependent. In the models of SLO, GLO and SMLO, this dependence of the width $\Gamma_{j}^{\alpha}$ is taken into account as in Fermi-liquid theory, i.e.:

\begin{equation}
\label{EQ_5}
\Gamma_{j}^{\alpha}=\Gamma_{j}^{\alpha}(\varepsilon_{\gamma},T)=
\Gamma_{j}^{\alpha}(\varepsilon_{\gamma},T=0)+\Delta\Gamma_{j}^{\alpha}(T).
\end{equation}

For GLO and SLO models:

\begin{equation}
\label{EQ_6}
\Delta \Gamma _{j}^{\alpha } (T)=g^{\alpha } \cdot 4\pi ^{2} T^{2} ,\, \, \, \, \, g^{\alpha } =\frac{\Gamma _{r,j}^{\alpha } }{(E_{r,j}^{\alpha } )^{2} } .
\end{equation}

The expressions (\ref{EQ_5}), (\ref{EQ_6}) at $\varepsilon_{\gamma}= E_{r,j}^{\alpha } $  correspond to the spreading width within framework of the Fermi-liquid theory in low-temperature limit \cite{Kol1995,Mug2000} with the normalization to the GDR width at zero temperature: $\Gamma _{r} (E_{r} ,T)=g\cdot (E_{r} ^{2} +4\pi ^{2} T^{2} )$, $g=\Gamma _{r} (T=0)/E_{r} ^{2} $. Note that, the temperature dependence of the experimental  GDR widths in warm nuclei can be described well by the expressions (\ref{EQ_4}), (\ref{EQ_5}) \cite{Den1989,Kol1995,Mug2000,Mug2002}. So, the total widths for SLO, GLO and SMLO models are given by the following expressions:  $\Gamma_{j}^{{\rm SLO}} (\varepsilon_{\gamma},T)=g^{{\rm SLO}} \cdot $$((E_{r,j}^{{\rm SLO}} )^{2} +4\pi ^{2} T^{2} ),$ $\Gamma _{j}^{{\rm GLO}} (\varepsilon _{\gamma } ,T)=g^{{\rm GLO}} \cdot $$(\varepsilon _{\gamma }^{2} +4\pi ^{2} T^{2} ),$  $\Gamma _{j}^{{\rm SMLO}} (\varepsilon _{\gamma } ,T)$$=g^{{\rm SMLO}} \cdot $ $(\varepsilon _{\gamma } \cdot E_{r,j}^{{\rm SMLO}} +4\pi ^{2} T^{2} ).$ The energy-dependent components of the widths correspond to transitions of the coherent 1p-1h states generating the GDR to the 2p-2h states and $\Delta \Gamma _{j}^{\alpha } (T)$ is connected with transitions to the 2p-2h states at thermo-dynamical equilibrium with the temperature $T$.

It should be mentioned that the temperature dependence for original TLO model was not presented, but we adopted the dependence like in the SLO model.

For GDR characteristics (the energies, widths and weights) in the SLO and SMLO models, the recommended values from recent data-base were used \cite{Plu2018}. The GDR parameters of SLO approach were taken for GLO model. The resonance energies and width of the TLO model were taken from ref.\cite{Jun2008}. The Bohr parametrization of the axis lengths was used for TLO model and the deformation parameters were taken from ref.\cite{Del2010}, where they were calculated within constrained Hartree-Fock-Bogoliubov approach with five-dimensional collective Hamiltonian.

The photodecay PSF for M1 transitions and corresponding input parameters were calculated with allowance for the scissors, spin-flip modes and upbend \cite{Gor2019,Gor2018}:

\begin{equation}
\label{EQ_7}
\begin{array}{c}
{\displaystyle
\overleftarrow{f}_{M1} (\varepsilon _{\gamma })= C\exp (-\eta \varepsilon _{\gamma })+} \\
{\displaystyle
+\frac{1}{3(\pi \hbar c)^{2} } \cdot \sigma _{sc} \cdot \frac{\varepsilon _{\gamma } \, \Gamma _{sc}^{2}}{\left({\varepsilon _{\gamma }^{2} -(E_{sc}^2)} \right)^{2} +\left({\Gamma _{sc} \varepsilon_{\gamma } }\right)^{2}} + } \\
{\displaystyle
+\frac{1}{3(\pi \hbar c)^{2} }\cdot \sigma _{sf} \cdot \frac{\varepsilon _{\gamma } \, \Gamma _{sf}^{2}}{\left({\varepsilon _{\gamma }^{2} -(E_{sf}^2)} \right)^{2} +\left({\Gamma _{sf} \varepsilon_{\gamma } }\right)^{2}} . } \\
\end{array}
\end{equation}

\begin{figure}  [b]
%    \centerline{\includegraphics[width=0.95\columnwidth]{Eps/Fig_1.eps}}
    \centerline{\includegraphics[width=0.95\columnwidth]{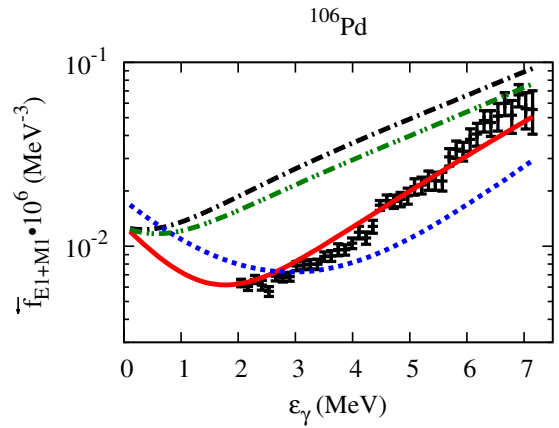}}    
	\caption{The photon strength function for gamma decay for nuclei ${}^{106}$Pd in comparison with calculations within different PSF models: $\bm{{\color{black} -\, \cdot \, -\, \cdot \, -}}$ SLO (black dash-dotted line),  $\bm{{\color{red}^{\_ \_ \_ \_ \_ \_ \_ } }}$ SMLO (solid red line)   ,  $\bm{{\color{blue} \cdot  \cdot  \cdot  \cdot  \cdot }}$ GLO (blue dotted  line), $\bm{{\color{green} -\, \cdot \, \cdot \, -\, \cdot \, \cdot \, -}}$ TLO (green dashed followed by two dots line). Experimental data are taken from ref.\cite{Oslo}.}
\label{fig_1}
\end{figure}

The sum of E1 and M1 PSF $\overleftarrow{f}_{E1+M1}^{\alpha}=\overleftarrow{f}_{E1}^{\alpha}+\overleftarrow{f}_{M1}^{\alpha}$ was calculated for gamma-decay analysis. For the comparison of the data two criteria were used: 1) minimum of the least-square value $\chi_{\alpha}^{2}$ and 2) minimum of the root-mean-square (rms) deviation factor  $f_{\alpha}$:

\begin{equation}
\label{EQ_8}
\begin{array}{c}
{\displaystyle\chi _{\alpha }^{2} =\frac{1}{n} \sum _{i=1}^{n}\frac{(\overleftarrow{f}_{exp} (\varepsilon _{i})-\overleftarrow{f}_{E1+M1}^{\alpha } (\varepsilon _{i}))^{2} }{(\Delta \overleftarrow{f}_{exp} (\varepsilon _{i}))^{2}}}; \\
\\
{\displaystyle f_{\alpha } =\exp \{ \chi _{\ln ,\alpha } \} ,} \\
\\
{\displaystyle \chi _{\ln ,\alpha }^{2} =
 \frac{1}{n} \sum _{i=1}^{n}\{ \ln \overleftarrow{f}_{E1+M1}^{\alpha } (\varepsilon _{i} )-\ln \overleftarrow{f}_{exp} (\varepsilon _{i} ) \} ^{2} }.
\end{array}
\end{equation}

Here $\overleftarrow{f}_{exp}$ is the experimental PSF data for even-even nuclei measured by OSLO group \cite{Oslo}, namely, for ${}^{44,46}$Ti, ${}^{56}$Fe, ${}^{92,94,96,98}$Mo, ${}^{106,108}$Pd, ${}^{106,112}$Cd, ${}^{116,122}$Sn, ${}^{148}$Sm, ${}^{162,164}$Dy, ${}^{166}$Er, ${}^{170,172}$Yb, ${}^{206,208}$Pb, ${}^{232}$Th, ${}^{238}$U.

The comparison of the different PSF models with the experimental data for ${}^{106}$Pd is  shown in fig.\ref{fig_1} as an example. The temperatures were calculated using the Fermi-gas model from EMPIRE code  \cite{EMPIRE}. It can be seen, that the use of SMLO model for E1 PSF describe rather well the experimental photodecay data within experimental uncertainties. The averaging of the theoretical PSF over the excitation energy \cite{Plu2012,Plu2014} only slightly changes the result.

\begin{figure} %[htbp]
    \includegraphics[width=0.95\columnwidth]{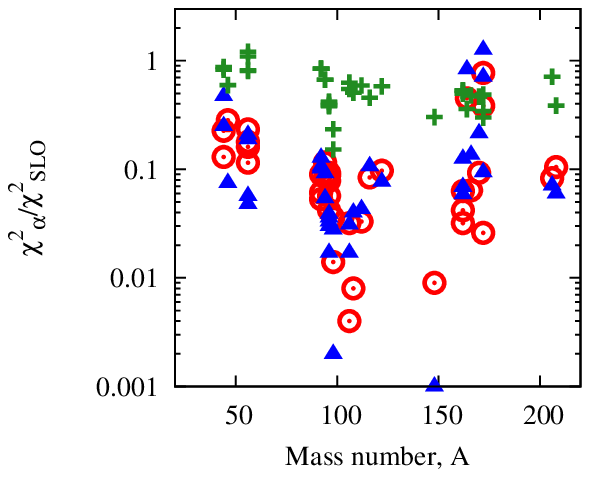}
    \includegraphics[width=0.95\columnwidth]{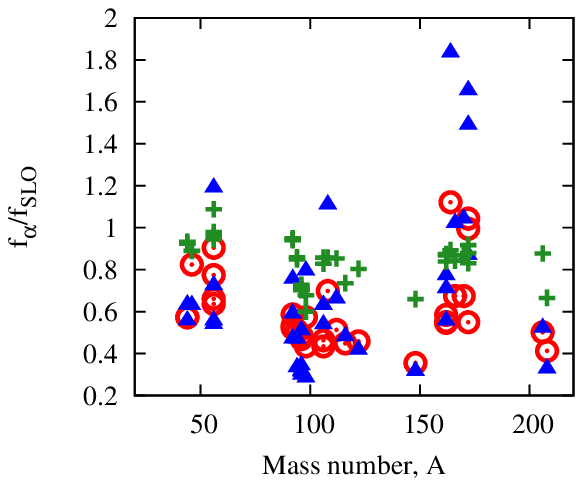}
	\caption{The relative values of rms deviation $\chi^2_{\alpha } /\chi^2_{{\rm SLO}} $ (upper figure) $f_{\alpha } /f_{{\rm SLO}} $ (bottom figure) for even-even isotopes corresponding to calculations within different PSF models. Signs:  red empty  circles ($\bm{{\color{red}\odot }}$)  -  SMLO;  blue empty triangles ($\bm{{\color{blue}\bigtriangleup }}$) -  GLO; green pluses (\textcolor{green}{\textbf{+}}) - TLO.}
\label{fig_2}
\end{figure}

\begin{table} %[htbp]
\center
\caption{{The ratios of mean values of least-square deviation $\chi^2_{\alpha}$ and mean values of rms deviation factor $f_{\alpha}$ for different E1 PSF models in relation to SLO.}}
\label{tabl1}
\begin{tabular}[c]{cccc}
\hline\noalign{\smallskip}
Criterion & SMLO & GLO & TLO\\
\noalign{\smallskip}\hline\noalign{\smallskip}
$<\chi^2_{\alpha}>/<\chi^2_{\rm SLO}>$ & 0.10& 0.04 & 0.40 \\
$<\chi^2_{\alpha}/\chi^2_{\rm SLO}>$ & 0.21& 0.29 & 0.71 \\
$<f_{\alpha}>/<f_{\rm SLO}>$ & 0.54& 0.58 & 0.78 \\
$<f_{\alpha}/f_{\rm SLO}>$ & 0.63& 0.81 & 0.84 \\
\noalign{\smallskip}\hline
\end{tabular}
\end{table}

The $\chi_{\alpha}^{2}$ and $f_{\alpha}$ criteria calculated for different PSF models are shown in the table \ref{tabl1} and fig.\ref{fig_2}. It should be mentioned that, according to the criteria of the minimum of these values, the $\overleftarrow{f}_{E1+M1}$ with the E1 SMLO model better describe the experimental data for the gamma-decay.

\section{Conclusions}
\label{sec:4}

The description of gamma-decay data from OSLO group by closed-form Lorentzian models of photon strength functions is considered. The experimental data are compared with theoretical calculations for even-even nuclei using criteria of minimum of least-square factor and root-mean-square deviation factor.
According to the comparison, one can conclude that as well as for the photoabsorption data \cite{Plu2018,Gor2019}, the SMLO model gives better description of the experimental data for gamma-decay and can be recommended in the nuclear reaction codes for modeling of the E1 photon strength function.

%\section*{Acknowledgments}
\begin{acknowledgement}
The authors are very thankful to Stephane Goriely and Tamas Belgya for valuable discussions, and Tamas Belgya for comments on TLO model. This work is partially supported by the International Atomic Energy Agency through a Coordinated Research Project on Updating the Photonuclear Data Library and generating a Reference Database for Photon Strength Functions (F41032).
\end{acknowledgement}

%
% BibTeX users please use
%\bibliographystyle{epj}
%\bibliography{activation_review}
%
% Non-BibTeX users please use

\end{sloppypar}
\end{document}